\documentclass[twocolumn,showpacs,amsmath,amssymb,floatfix,aip,jcp,reprint]{revtex4-1}
\pdfoutput=1
\usepackage{graphicx,color,subfigure} % use for side-by-side figures
\usepackage{bm,bbold,empheq,mathdots}
\usepackage{units,enumitem}
\usepackage{hyperref} % use for hypertext links, including those to external documents and URLs
 \newcommand{\be}{\begin{equation}}
 \newcommand{\ee}{\end{equation}}

\newcommand{\e}{{\text{e}}}
 \renewcommand{\d}{{\text{d}}}
 
\DeclareMathAlphabet{\mathvar}{OT1}{pzc}{m}{it} %for q in Zapf Chancery

  \newcommand{\xref}[1]{(\ref{#1})}
  
\begin{document}
	\title{Phase diagram of selectively cross-linked block copolymers
	%: Route to a			%		lamellar %
	shows chemically %compositionally 
	microstructured
	gel}
	
	\author{Alice von der Heydt}\email{heydt@theorie.physik.uni-goettingen.de}
	\affiliation{Institut f\"ur Theoretische Physik, Georg-August-Universit\"at G\"ottingen, 		Friedrich-Hund-Platz 1, 37077 G\"ottingen, Germany}

	\author{Annette Zippelius}
	\affiliation{Institut f\"ur Theoretische Physik, Georg-August-Universit\"at G\"ottingen,
	Friedrich-Hund-Platz 1, 37077 G\"ottingen, Germany}
	\affiliation{Max-Planck-Institut f\"ur Dynamik und {Selbst}\-organisation, 
	Am Fa{\ss}berg 17, 
	37077 G\"ottingen, Germany}

\date{\today}

\begin{abstract}
We study analytically the intricate phase behavior of cross-linked $AB$ diblock copolymer melts, which can undergo two main phase transitions due %owing 
to %additional 
quenched random constraints: 
Gelation, \textit{i.e.}, spatially random 
localization of polymers forming a system-spanning cluster,
%(amorphous solidification or gelation) 
is driven by increasing the number parameter $\mu$
of irreversible, type-selective cross-links between random pairs of $A$ blocks. 
Self-assembly %ordering
into a periodic pattern of $A$/$B$-rich microdomains (microphase separation) 
is controlled by the $AB$ incompatibility $\chi$ inversely proportional to temperature.
Our model aims to capture the system's essential microscopic features, 
including an ensemble of random networks %realizations 
that reflects spatial correlations at the instant of cross-linking.
We identify suitable order parameters 
and derive a free-energy functional %governing the interaction fields.
%An expansion of this functional 
in the spirit of Landau theory
that allows us to 
trace a phase diagram
in the plane of $\mu$ and $\chi$.
%Lowering the temperature in the mixed liquid state, 
Selective cross-links promote microphase separation at higher critical temperatures
%incompatibilities $\chi_c$ smaller 
than in uncross-linked diblock copolymer melts.
%, in agreement with intuition and earlier studies.
%Increasing the cross-link density in the mixed liquid state reproduces the known transition to an isotropic gel at a critical mean polymer coordination $\mu_c=1$.
Microphase separation in the liquid state facilitates %prepares 
gelation, giving rise to 
%so that at $\mu_c < 1$, 
a novel gel state whose chemical composition density mirrors the 
periodic $AB$ pattern. 
%imprinted in chemical composition
%with periodically varying $A$ density along the lamellar domain normal emerges. 
%which is expected to be %frozen-in
%stabilized by cross-links against mixing upon %subsequent 
%heating.
\end{abstract}

 %Valid PACS appear here
\maketitle

\section{Introduction}

Block copolymer melts are known to self-assemble into a
variety of complex ordered microstructures
\cite{leibler80,bates-fred90rev,matsenPRL94}.
Combining block copolymers and quenched disorder is considered promising, \textit{e.g.}, 
for the design of biomimetic materials with tunable micropatterns \cite{chakrab2001}.
Irreversible, random cross-linking provides one method to introduce quenched disorder.
On the one hand, sufficient cross-linking can stabilise 
the so-called microphases over a
wide range of temperatures. 
%Alternatively, 
On the other hand, cross-links in a chemically homogeneous gel
can prevent ordered microphase separation, 
similarly as cross-links hinder macroscopic phase
separation in a blend of homopolymers \cite{sfatos97rev,wald-zipp-goldb-epl05,wald-goldb-zipp06}. 
In general, there is a competition between ordering and random topological constraints
due to cross-linking.
\textit{Selective} cross-links in biological heteropolymers provide cells with intriguing mechanisms of adaption: One example is
%In nature, selective cross-links are present, \textit{e.g.}, in 
peptidoglycan, also called murein, 
which is composed of sugars and amino-acids
and forms a mesh-like wall around many bacteria's plasma membrane \cite{burge77,rogers80book}.
%form outer cell wall
%selective cross-links between sugars form mesh structure

For randomly and irreversibly cross-linked block copolymer melts, 
two phase-ordering mechanisms can be controlled \textit{independently}: 
The incompatibility between the two chemical %ly different
components of the block copolymer,
usually quantified by the Flory-Huggins parameter $\chi$,
controls the ordering transition in local 
chemical composition %component concentrations %of the fluid
from a disordered (mixed) state to a periodic microstructure.
%as regards the densities of the chemical components. 
The number, or the chemical potential, $\mu$, of random, irreversible, 
and here component-selective, cross-links 
controls gelation, %giving rise to 
a transition from a fluid (sol) to an amorphous solid (gel) \cite{gcz96}. 
This equilibrium state without long-range order,
in which a fraction of polymers is localised at random positions,
has been the subject of continued interest \cite{panyuk-rabin96}. 

Once the disordered sol state has undergone gelation, random localisation and
irreversibility of the cross-links rule out the formation of an ordered microstructure.
%and hence can no longer order into a microstructured state.
However, increasing the number of component-selective cross-links 
in the chemically ordered sol
allows for a novel phase state --- a microstructured gel. 
%In contrast to an isotropic, homogeneous gel,
%this phase exhibits a periodic microstructure with respect to
%%as far as the 
%chemical composition. %is concerned. 
%the independent control of the two phase-transition mechanisms combined with 
%microstructure in the gel is enabled by
%%%%%%%%%%%%%%%%
\begin{figure}
\centering
\includegraphics[width=8cm]{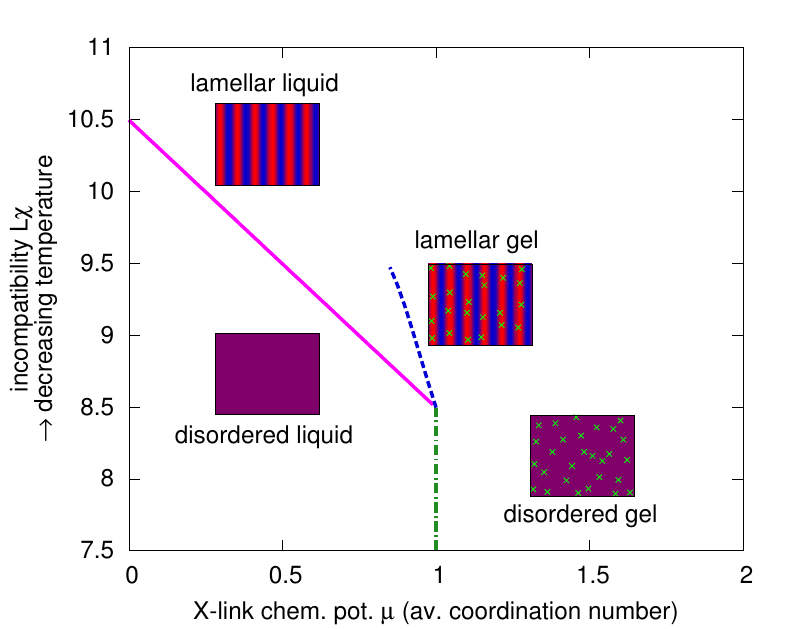}
\caption{\label{fig:phased} Phase diagram in the plane of incompatibility, $\chi$,
  and chemical potential for the
  cross-links, $\mu$; 
  four different phases are predicted: isotropic fluid,
  lamellar fluid, isotropic gel and lamellar gel.}
\end{figure}
%%%%%%%%%%%%%%%%%
This microstructured gel is anisotropic and heterogeneous in chemical composition:
In the simplest case of lamellae, 
segments of chemical component $A$ are localised randomly 
within each planar $A$-rich domain, 
but their concentration follows the periodic modulation of the lamellae in the third space dimension. 
The ordering is reminiscent of that
in a smectic liquid crystal, except for the fact that
within the planes, the particles in the gel cluster are not mobile as in a fluid but
randomly localised as in an amorphous solid. 
The complete phase diagram, 
as obtained by the approach presented in the following,
is
shown %schematically 
in Fig.~\ref{fig:phased} in the plane of
incompatibility, $\chi$, and cross-link chemical potential, $\mu$. 
Four different phases are present: an isotropic fluid, a lamellar (in
general, microstructured) fluid, an isotropic gel and a microstructured gel. 
As alluded to above, the irreversible cross-linking considered here 
impedes transitions
between the two gel phases. 
Which of the gel phases is actually
observed, depends on the state of the melt at the instant of cross-linking.

In this paper, we discuss a mean-field theory in terms of two order-parameter fields: 
one for phase separation and one for random
localisation. In this theory, the free energy takes the form of a Landau-Wilson
expansion, which can be derived from a microscopic
model. Alternatively, the Landau-Wilson free energy can be derived
phenomenologically from symmetry arguments. 
We discuss both approaches, deferring most of the
technical calculations to appendices. First, in Sec.~\ref{sec:model}, 
we set up a microscopic model for randomly cross-linked block
copolymers. Then, in Sec.~\ref{sec:orderpar}, we go on to discuss the collective fields
or order parameters which follow naturally from the microscopic model
and have an intuitive interpretation in terms of microphase ordering and
gelation. In Sec.~\ref{sec:free-energy}, we present the Landau-Wilson free energy
used in the subsequent sections to obtain the phase diagram.
In Sec.~\ref{sec:lam-fluid}, we discuss the lamellar fluid, 
and in Sec.~\ref{sec:lam-gel}, the instability to the lamellar gel. 
Conclusions and an outlook are presented in Sec.~\ref{sec:conclusion}.

\section{\label{sec:model}Model}
\subsection{\label{subsec:diblocks}Diblock copolymer melt}

The melt comprises $N$ monodisperse diblock copolymers with degree of polymerization $L$ confined to a volume $\tilde V$. The global segment density is $\tilde\varrho_0=NL/\tilde V$. 
We approximate the polymer contours $\bm R_j(s)$, $j=1,\ldots,N$, $s\in[0,1]$, 
as Gaussian chains with radius of gyration $R_G = \sqrt{Lb^2/6}$ in three dimensions 
($b$ is the size of one statistical segment).
For convenience, we choose units of energy such that $k_BT=1$ 
and measure all lengths in units of $R_G$,
so that dimensionless Cartesian coordinates $r_{j, \nu}$ are defined according to 
$r_{j, \nu} = R_{j, \nu}/R_G$. %($\nu$ indexes Cartesian components).
The dimensionless system volume is $V = \tilde V /R_G^3$, 
and the  global (dimensionless) segment density is denoted by $\varrho_0$.
Accordingly, polymer conformations are governed by a quadratic effective
potential
\be 
\mathcal{H}_{N\text G}  = 
\frac{1}{4} \sum_{j=1}^{N} \int_0^1\! \d s
\left| \partial_s \bm r_j(s)\right|^2, 
\ee
and in the following, $\mathcal{H}_{\text G}$ is used to denote the single-chain potential.

%\subsection{Excluded volume interaction}
For the inter-polymer interactions whose definitions follow, 
we adopt a semi-microscopic %coarse-grained 
view, in which 
interaction shapes can be approximated by delta functions.

All polymer segments irrespective of their type interact via the excluded volume repulsion, 
which is of chiefly entropic origin and thus can be written as
\be\label{H-ev}
\mathcal{H}_{\text{ev}}  = 
 \frac{ \kappa NL}{4\varrho_0^2} 
\int \d^3 r 
\varrho(\bm r)  \varrho(\bm r)
\ee
where $\kappa > 0$ is the (dimensionless) compression modulus, 
and $\varrho(\bm r) \mathrel{\mathop:}= 
\frac{L}{V} \sum_{j=1}^{N} \int_0^1 \!\d s\, \delta(\bm r - \bm r_j(s)) $ is 
the total density.
%\subsection{Incompatibility}

Since the van-der-Waals attraction between chemically different polymer segments is weaker than between equal ones, there is a net repulsion or incompatibility between type $A$ and $B$.
In each diblock copolymer, the block of type $A$ comprises a fraction $f\in[0,1]$ of the chain length,
%(the total number of segments), 
and the block of type $B$ a fraction $1-f$.
A binary variable $q(s)$ records the type at contour position $s$ via
\be
\label{type-variable}
q(s) \mathrel{\mathop:}= \left\{
\begin{array}{ll}
+1, & s \text{ is of type } A,\\
-1, & s \text{ is of type } B
\end{array}
\right.
\ee
(due to monodispersity, $q(s)$ does not depend on the polymer index $j$).
The average value $\mathvar q = 2f-1$, related to the $A$ fraction $f$, 
quantifies the global excess of $A$ segments. 
Expressed by 
the local imbalance or $A$ excess density %of $A$ segments
\be
\sigma(\bm r) \mathrel{\mathop:}= 
\frac{L}{V} \sum_{j=1}^{N} \int_0^1 \!\d s \left[q(s) - \mathvar{q}\right] 
\delta\left( \bm r - \bm r_j(s) \right),
\ee
the incompatibility interaction between $A$ and $B$ segments takes the form
\be\label{H-chi}
\mathcal{H}_{\chi}  = 
- \frac{ \chi NL}{4\varrho_0^2} 
\int \d^3 r \sigma(\bm r) \sigma(\bm r)
\ee 
where $\chi$ denotes the incompatibility (Flory) parameter, which is assumed to be positive.

\subsection{Random, irreversible, type-selective cross-links}
%%%%%%%%

\begin{figure}
\centering
\includegraphics[width=.4\columnwidth]{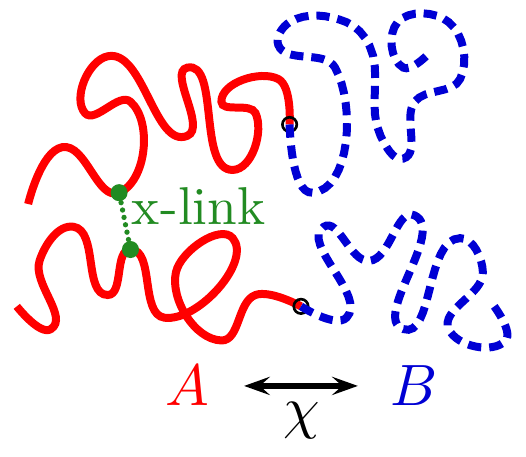}
\includegraphics[width=.4\columnwidth]{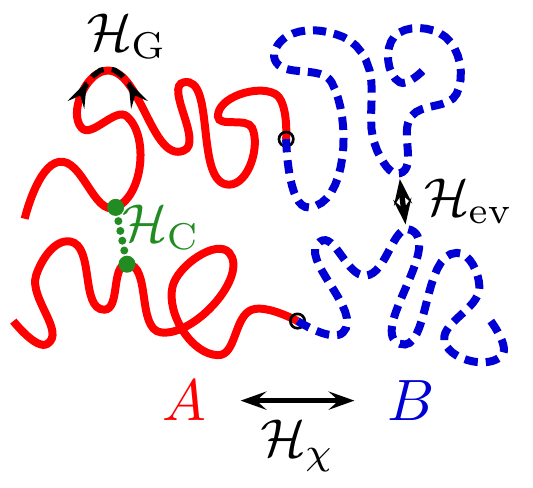}
\caption{\label{fig:x-link}Selective cross-link between the $A$ blocks of two diblock copolymers and interactions considered in the model.}
\end{figure}

Motivated by experiment \cite{Tietz14}, we consider a synthesis scheme, 
which incorporates cross-linkers \textit{selectively}, here only into the
$A$ blocks of the copolymers, cf.\ Fig.~\ref{fig:x-link}. 
%%%%%%%%%%
Randomly chosen pairs of $A$ blocks, $j_m, j'_m \in \{1,2,\dots,N\}$, are permanently cross-linked at random arc-lengths, $s_m, s'_m\in [0,f]$,
so that one configuration %realization 
(instance) of $M$ cross-links is specified by the set of pairs
${\cal C}\mathrel{\mathop:}=\left\{ \left( (j_m,s_m), (j'_m, s'_m) \right) \right\}_{m=1}^M$.
The cross-links are modeled as entropic, harmonic springs of (tunable) constant $1/(2a_c^2)$,
\be\label{H-Xlink}
\mathcal{H}_{c}  = 
\frac{1}{4 a_c^2} \sum_{m=1}^M \left| \bm r_{j_m}(s_m) - \bm r_{j'_m}(s'_m)\right|^2.
\ee

Pursuing a statistical mechanical approach, 
we aim at computing the free energy of the model specified by the total Hamiltonian
\be\label{H-total}
\mathcal{H}\left\{q(s), \bm r_j(s) \right\}=\mathcal{H}_{N\text G}+\mathcal{H}_{\text{ev}}+\mathcal{H}_{\chi}
+\mathcal{H}_{c},
\ee
the parts of which are sketched in the right panel of Fig.~\ref{fig:x-link}.
Rather than in the absolute free energy, we are interested in the free-energy \emph{difference\/}
between the cross-linked diblock copolymer melt with $AB$ incompatibility
and a disordered, single-type 
(homo)polymer melt with uniform densities and with no interactions but intra-chain connectivity.
Therefore, we normalise the canonical partition functions
with the partition function of a homogeneous melt of $N$ non-interacting Gaussian chains, %without $AB$ repulsion,
\be
{\cal{Z}}({\cal C}) = \frac{
\int\!{\cal D}\left[\bm r_j(s)\right] \, \e^{ -
\displaystyle{ 
\mathcal{H} \left\{q(s), \bm r_j(s) \right\} }
} 
}{ 
\int\!{\cal D}\left[\bm r_j(s) \right]\, \e^{ -\displaystyle{{\cal H}_{N\text{G}} 
\left\{\bm r_j(s)\right\}} }  
}.
\label{norm-part-fct}
\ee
Herein, ${\cal D}\left[\bm r_j(s)\right]$ denotes the measure of 
functional integrations over the polymer configurations $\bm r_j(s)$, $j=1,\dots, N$, $s\in[0,1]$. 
%continuous-chain polymers
%(the $NLd$-fold integration over the segment coordinates for discrete Gaussian chains).
Since we consider a random ensemble of \textit{irreversible} cross-links,
the disorder due to cross-links is quenched, 
hence the quantity to be reasonably disorder-averaged 
is the free energy, \textit{i.e.}, $-\ln {\cal{Z}}({\cal C})$, 
where ${\cal{Z}}({\cal C})$ is the canonical partition function of cross-link configuration $\cal C$.

We choose a grand-canonical cross-link ensemble, in which instead of
the number of cross-links, $M$, their chemical potential $\mu > 0$ is
fixed. All cross-links are assumed to be formed simultaneously and
instantaneously in the liquid phase 
--- a process which is well modeled
by the Deam-Edwards distribution~\cite{Deam-Edwards} 
\be\label{DE-distrib}
\mathcal{P}_{\text{DE}}({\cal C};M ) \propto \mathcal{Z} ({\cal C})
\frac{\mu^M}{M!}.  
\ee 
From the cross-link chemical potential $\mu$, the average polymer coordination number in the network follows as
$\mu = 2 \left[ M \right]_{\mathcal{P}_{\text{DE}}} /N$.
Via including the partition function at preparation,
the Deam-Edwards distribution `measures' the loss in 
configurational entropy due to the cross-link constraints and thereby
accounts for spatial correlations %at the instant of cross-linking.
in the uncross-linked melt. 
Also, the distribution allows to choose
the preparation state, in which cross-linking is performed,
to be different from the measurement state, \textit{i.e.}, 
the state in which the network is probed in experiment. 
Here, for a first overview, we assume 
the preparation and the measurement state to be the same. 

With the cross-link distribution of Eq.~\xref{DE-distrib}, 
the specification of the model is complete, 
and we can proceed to derive the disorder-averaged free energy from first principles. 
The resulting saddle-point equations 
for the order parameters for phase separation (ordering), gelation, and possibly 
mixed transitions, then have to be solved self-consistently. 
The basic steps of this procedure, 
eventually yielding a free energy of Ginzburg-Landau-Wilson form,
are sketched in the appendix. 
In the main text, we will consider this Ginzburg-Landau-Wilson free energy only, 
since its essential
features can be derived from general symmetry principles, without the use of replicas. 
This requires a careful discussion of the order parameters, 
which is the focus of the next section.

\section{\label{sec:orderpar}Order Parameters}
%We expect to find 4 different phases:
%\begin{itemize}
%\item{isotropic liquid}
%\item {lamellar liquid}
%\item{isotropic gel}
%\item{lamellar gel}
%\end{itemize}
%We discuss the appropriate order parameters in the order of increasing complexity. 

\begin{figure}
\centering
\includegraphics[width=8cm]{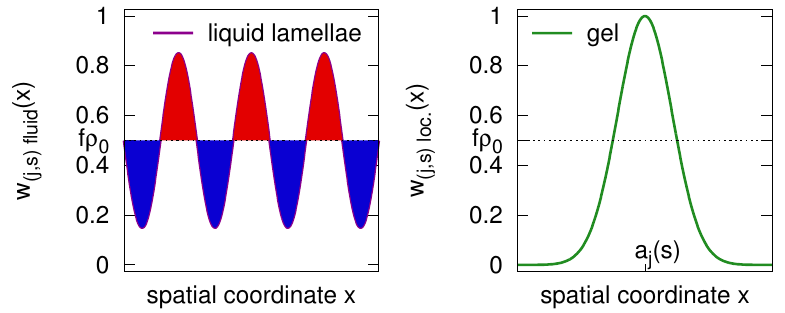}
\caption{\label{fig:probdens}One-dimensional, single-$A$ probability density distribution in the fluid lamellar (left) and in the %isotropic 
gel (right) state.}
\end{figure}

Phase separation into %periodically ordered %arranged 
$A$/$B$-rich domains is signaled 
by spatial modulations of the imbalance density
\be
\label{order_par_AB}
\sigma(\bm k) 
 = 
 \frac{L}{V} \sum_{j=1}^N\int_0^1\! \d s 
\, \left( q(s)-\mathvar q \right) \left\langle \e^{i\bm k \cdot \bm r_{j}(s)} \right\rangle,
\ee
cf.\ Eq.~\xref{type-variable} for the type variable $q(s)$,
where $\langle\cdot\rangle$ denotes the thermal average with the total Hamiltonian from Eq.~\xref{H-total} for a fixed cross-link configuration.
Here, we will mainly address symmetric $AB$ diblock copolymers,
\textit{i.e.}, $A$-fraction $f=1/2$ or $\mathvar{q} = 0$, 
in which a lamellar state arises, the prototype of an ordered microstructure. 
The imbalance density in this state can be parametrized 
by a %periodic
one-dimensional, single-mode distribution in space,
\be\label{probdensity_lam}
\sigma({\bm x}) % = \sum_{j} \int_0^1\!\d s q(s) \left\langle \delta ({\bm x}-\bm r_{j}(s))\right\rangle
=\sigma_0 \cos\left( \bm q_0 \cdot \bm x + \phi \right),
\ee
characterized by the amplitude $\sigma_0$, the wavevector ${\bm q}_0$ and a phase $\phi$. 
In fact, there is a manifold of symmetry-related states defined by $\phi$ and the direction of $\bm q_0$, 
and we are free to choose one of them, viz., $\phi=0$ and $\bm q_0 = q_0 \bm e_1$. 
The corresponding probability density distribution $w_{(j,s)\, \text{fluid}}({\bm x})=\bigl\langle
\delta ({\bm x}-\bm r_j(s))\bigr\rangle$ of a single mobile (fluid) $A$ segment is sketched 
in the left panel of Fig.~\ref{fig:probdens}.

An $A$ segment $(j,s)$ 
in the gel state is localised at a random position ${\bm a}_j(s)$.
In a Gaussian localisation model, the probability distribution is sharply peaked around ${\bm a}_j(s)$
\be
w_{(j,s)\, \text{loc.}}({\bm x})=\bigl\langle
\delta ({\bm x}-\bm r_j(s))\bigr\rangle
\propto \e^{ - \left( \bm x - \bm a_j(s)\right)^2/(2 \xi^2) },
\ee
as shown schematically in the right panel of Fig.~\ref{fig:probdens}. 
The extent of spatial localisation is measured by the localisation length $\xi$.
Since the positions are random, the average density vanishes in the thermodynamic limit. 
The macroscopic observable
able to detect random localisation with a minimal number of arguments 
is %the normalized $A$ overlap density
\be
\label{OP2replica}
\Omega( \bm k, \bm p ) = \frac{1}{Nf} \sum_{j=1}^N\int_0^f \!\!\d s 
\left\langle\e^{i\bm k\cdot\bm r_{j}(s)}\right\rangle \left\langle\e^{i\bm p\cdot\bm r_{j}(s)}\right\rangle.
\ee
Due to the $A$-selective cross-links or topological constraints, it suffices to sum over the $A$ segments.
%Here, we take into account the $A$-monomers only, because only these are cross-linked. 

Since phase separation and gelation are controlled independently, we
expect to find four different phases:
\begin{center}
\begin{tabular}{lcl}
\textbullet\, isotropic liquid & & \textbullet\; isotropic gel\\
\textbullet\, lamellar liquid & & \textbullet\; lamellar gel
\end{tabular}
\end{center}
%\begin{itemize}
%\item{isotropic liquid}
%\item {lamellar liquid}
%\item{isotropic gel}
%\item{lamellar gel}
%\end{itemize}

In the isotropic fluid, both order parameters vanish. 

In the isotropic gel, the Gaussian model \cite{gcz96} predicts
 \begin{align}\label{orderpargel}
  \sigma(\bm k)&=0 \text{ and} \\
  \Omega_{\text{iso}}({\bm k},{\bm p}) &= 
 \underbrace{ 
 Q\, \delta_{\bm k+\bm p,0} \,e^{-k^2 \xi^2/2}
 }_{\delta \Omega_{\text{iso}}(\bm k, \bm p)}
  + \,(1-Q)\, \delta_{\bm k,0} \delta_{\bm p, 0} .
\end{align}
The finite fraction of localised $A$-blocks is denoted by $Q$,
and macroscopic translational invariance of the amorphous gel
requires $\bm k+\bm p=0$ for the gel part $\delta \Omega_{\text{iso}}$.

In the  \emph{lamellar} fluid, both order parameters are nonzero, 
even though the particles are
not localised in the lamellar fluid, but exhibit periodic density
modulations.
Consistent with Eqs.~\xref{probdensity_lam} and \xref{OP2replica},
we can deduce the following minimal parametrizations
of the order parameters in the lamellar fluid:
\begin{align}
  \sigma_{\text{lam}}({\bm k})&=
  \sigma_0\;(\delta_{\bm k,\bm q_0} + 
\delta_{\bm k,-\bm q_0}), \\
  \Omega_{\text{lam}}({\bm k},{\bm p})&\propto\sigma_0^2 
(\delta_{\bm k,\bm q_0} + \delta_{\bm k,-\bm q_0})
(\delta_{\bm p,\bm q_0} + \delta_{\bm p,-\bm q_0}).
\label{orderparlam}
\end{align}

In the \emph{lamellar gel\/}, we expect the order parameter Eq.~\xref{OP2replica} to reflect both ordering and localisation: 
There is the lamellar-fluid contribution, Eq.~\xref{orderparlam}, 
and a true gel contribution $\delta\Omega({\bm k},{\bm p})$, 
which accounts for localisation of a fraction $Q$ of polymers.
%$\Omega({\bm k},{\bm p})= \Omega_{\text{lam}}({\bm k},{\bm p}) + \delta\Omega({\bm k},{\bm p})$. 
To reflect the lamellar order imprinted in the gel, $\delta\Omega$ must again split up and feature, in addition to the isotropic gel $\delta \Omega_{\text{iso}}$, Eq.~\xref{orderpargel}, a novel part.
Inserting the Gaussian localisation model, we find %for $\delta\Omega$
\be
\delta\Omega({\bm k},{\bm p})=% \frac{1}{V} \sum_{j=1}^N\int_0^f ds 
%<\e^{i\vec k\cdot\vec r_{j}(s)}><\e^{i\vec p\cdot\vec r_{j}(s)}>=
\frac{Q}{Nf} \!\sum_{j=1}^N\!\int_0^f \!\! \d s\, 
\e^{i(\bm k+\bm p)\cdot\bm a_{j}(s)} \e^{-(k^2+p^2)\xi^2/2}.
\ee
To perform the spatial %network realization 
average 
over (partially) random localisation centres $\{\bm a_{j}(s)\}$ of $A$, 
we need to know their probability distribution. % $p(\{\bm a_{j}(s)\} )$.
In a mean-field picture,
we can assume that all $\bm a_j(s)$ 
are independently, identically distributed.
The single-centre probability distribution %depends on $\bm a$ only:
\be
p(\bm a) %=\overline{\delta(\bm x-\bm a)}
= \frac{1}{V} %f \varrho_0 
+ \frac{\sigma_0}{\varrho_0}  
\cos\left( \bm q_0 \cdot \bm a \right)
\ee 
displays a uniform part as in the isotropic gel state 
and a periodic modulation in the direction of $\bm q_0$,
generated by the pattern of $A$-rich microdomains and 
proportional to $\sigma_0$ (\textit{cf.}, the single-$A$ density distribution in the left panel of Fig.~\ref{fig:probdens}, %extra Fig.? \ref{fig:lam-gel-distrib}.
%The difference is 
but recall that $p(\bm a)$ is the mean-field distribution of a \emph{localisation} position.)
With this distribution, the lamellar gel part of
the order parameter becomes
\begin{align}
\lefteqn{
\delta\Omega({\bm k},{\bm p})}\label{orderparlam-gel}\\
& = Q\,
\big(
\delta_{\bm k+\bm p, \bm 0}
+ \frac{\sigma_0}{\varrho_0} \left(
\delta_{\bm k+\bm p, \bm q_0}+\delta_{\bm k+\bm p, -\bm q_0} \right)
\bigr) 
\e^{-(k^2+p^2) \xi^2/2}.
\notag\end{align}
A more elaborate theory should account for an anisotropic degree of localisation, 
\textit{i.e.}, allow for different  localisation lengths 
within the plane of a lamellar domain
and perpendicular to it. %(larger than?) 
Here, to discuss the instabilities toward %and the general form of 
the different phases, there is no need to detail the form of localisation, %anyway 
so we leave this refined ansatz for future work. 
In the next section, we discuss the Landau-Wilson free energy in terms of the two order parameters introduced above, $\sigma(\bm k)$ and $\Omega({\bm k},{\bm p})$.

\section{\label{sec:free-energy}Landau-Wilson free energy}

As mentioned above, the Landau-Wilson free energy can be derived completely
from the microscopic model, and we summarize
this calculation in Appendix \ref{app:free-energy}. 
In this part, we just quote the result which 
--- apart from the precise form of the vertices --- 
follows also from symmetry principles.
As needed, we use %partially
symbolic notation for the vertices.
Also for reasons of transparency, we restrict the consideration to the $A\leftrightharpoons B$ exchange-symmetric diblock melt, \textit{i.e.}, $A$-fraction $f=1/2$.
%E.g., the elimination of total density fluctuations at the saddle point
%%in the 1RS
%does not affect the instability towards phase separation.

We decompose the total free energy into three parts:
\be
\label{Ftotal}
F \left[ \sigma,\Omega\right] =
F_{\sigma}[\sigma]+F_{\Omega}[\Omega]+
F_{\sigma\Omega}[\sigma,\Omega].
\ee

\begin{figure}
\includegraphics[width=5cm]{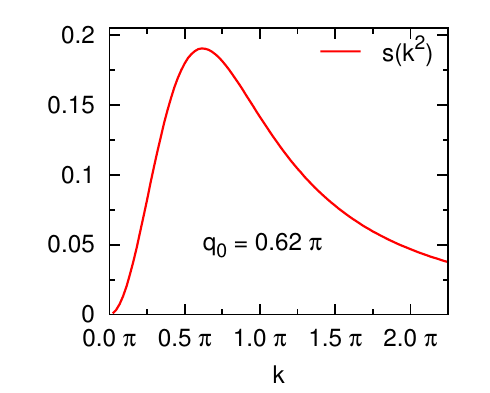}
\caption{\label{fig:s-k}%Quadratic $\sigma_{\bm k}$-
Correlation $s(k^2)$ for symmetric diblocks. 
%With restored units, the maximum occurs at a wave number of $\tilde q_0 = 0.62\pi/R_G$.
}
\end{figure}

The first one, $F_{\sigma}$,
accounts for lamellar phase separation and 
displays the standard form for symmetric $AB$ block copolymers (here, we abbreviate
$\sigma(\bm k)$ by $\sigma_{\bm k} $):
\begin{align}
F_{\sigma}[\sigma]  =  
& \frac{L\chi_{\mu}}{4\varrho_0^2}  \sum_{\bm k\neq \bm 0}  
 \left( 1 - L\chi_{\mu} \frac{s(k^2)}{2} \right) 
\sigma_{\bm k} \sigma_{-\bm k}
\label{h-sigma}\\
& + \left[\frac{L\chi_{\mu}}{\varrho_0}\right]^4 
\!\!\!\sideset{}{'}\sum_{\bm k_1, \bm k_2, \bm k_3
}\!\! s_{\sigma^4} \left( \{\bm k_l\}, \{\sigma_{\bm k_l}, \sigma_{- \sum_{l=1}^3 \bm k_l}\} \right)
%(\bm k_1,\bm k_2,\bm k_3)
%\sigma_{\bm k_1}\sigma_{\bm k_2}\sigma_{\bm k_3}\sigma_{- \sum_{l=1}^3 \bm k_l}
\notag
\end{align}
with the cross-link-rescaled incompatibility
\be\label{Lchimu}
L\chi_{\mu}  %^{(\alpha)}
\mathrel{\mathop:} = L\chi + 2 \mu. %\frac{\mu}{2f^2}
\ee
The correlation function in the quadratic term is
\begin{align}
%\lefteqn{
s(k^2)  %}\notag\\
& = \iint_0^1\!\d s_1 \d s_2\, q(s_1) q(s_2) 
\left\langle \e^{-i \bm k\cdot(\bm r (s_1) - \bm r(s_2)) } \right\rangle_{\!\mathcal{H}_{\text{G}} }
\notag\\
& = 4\, d_{1/2}(k^2) - d_1(k^2), 
\label{s_symm}\end{align}
with the Debye function
$d_a(k^2)  \mathrel{\mathop:}= 2 (\e^{-ak^2 } - 1 + ak^2)/k^4$. 
It exhibits the known feature of a maximum at a finite wavenumber $
q_0$, \textit{cf.}, Fig.~\ref{fig:s-k}, corresponding to a lamellar wavelength of $3.2 R_G$ at the onset of microphase separation.
%%%%%%%%%%%%%%%%%%%
%%%%%%%%%%%%%%%%%%%
Thereby, $q_0$ is the instability mode, since
upon increasing $\chi$, 
the homogeneously mixed phase with $\sigma_{\bm k}=0$ 
becomes unstable
%towards lamellae  %mode $q_0$
when the first coefficient of the
quadratic term (in the mixed state positive) vanishes: 
$2 - L\chi_{\mu,c} s(q_0^2)=0$.  
At this point,
cross-links influence the $AB$ order only via the rescaled incompatibility Eq.~\xref{Lchimu}:
For the unconstrained melt of diblocks, 
microphase separation occurs at $L\chi_c=2/s(q_0^2)$,
whereas selective cross-linking shifts this transition to lower $\chi$
(or higher temperatures): 
\be L\chi_c = \frac{2}{s(q_0^2)} - 2\mu
\ee 
This effect is known \cite{gomez-balsara06,lay-somm2000,lay-somm-blum99} and intuitively comprehensible: %understood:
Selective cross-links
give rise to %the formation of $A$-rich regions, favouring 
chemical ordering via the formation of $A$-rich regions and thus favour
$AB$ phase separation. 
Finally, the vertices of the fourth-order term, summarized by $s_{\sigma^4}
%(\bm k_1,\bm k_2,\bm k_3)
$ in the constrained sum $\sum'$, can be computed as correlation functions of the Gaussian theory explicitly defined in Appendix~\ref{app:vertices}. 

The second term in Eq.~\xref{Ftotal}, accounts for the gel transition in
the absence of phase separation. 
We restrict ourselves here to the
minimal complexity of the order parameter, $\Omega({\bm k},{\bm p})$, 
defined as the
second moment of the local density in Eq.~\xref{OP2replica}. 
A complete theory of
gelation requires all moments %, see, \textit{e.g.}, 
\cite{gcz96}. 
However, we will see shortly
that only the second moment couples to the order parameter for phase
separation, so that we can simplify the free energy considerably:
\begin{align}
\lefteqn{ F_{\Omega}[\Omega] =} \label{f_omega}\\
&\frac{\mu}{2}\sum_{\bm k,\bm p\neq 0}
\big(1 - 4 \mu d_{1/2}(k^2+p^2)\big)
\Omega({\bm k},{\bm p})\Omega({-\bm k},{-\bm p})\notag\\
&\mbox{} - \frac{\mu^3}{6}
\sum_{\bm k_1,\bm k_2\neq 0}
\sum_{\bm p_1,\bm p_2\neq 0}
s_{\Omega^3}(\bm k_1, \bm k_2; \bm p_1, \bm p_2)
\notag\\
& \quad\mbox{}\times
\Omega({\bm k_1},{\bm k_2})
\Omega({\bm p_1},{\bm p_2})\Omega({-\bm k_1-\bm p_1},{-\bm k_2-\bm p_2})
\notag\end{align}
Here, $s_{\Omega^3}$ is a vertex function (or constant, if evaluated at wave numbers zero), 
which can be computed from the microscopic model.  
The fluid phase (sol) with $\Omega=0$ becomes unstable to the
formation of an isotropic gel when the concentration of cross-links is
sufficiently high so that $1-\mu d_{1/2}(k^2+p^2)=0$. 
The maximum of the correlation function $d_{1/2}(k^2+p^2)$ occurs at $k^2+p^2=0$, 
so that without lamellar order, the gel transition occurs at $\mu_c=1$, 
independent of the incompatibility $\chi$.
This transition is marked by the vertical line in Fig.~\ref{fig:phased}.
The continuous growth of the gel fraction $Q$ as a function of $\mu - \mu_c$ 
is known already from the percolation description by Flory and Stockmayer \cite{flory41,stockm43}
and
can actually be computed by inserting the isotropic-gel ansatz for $\Omega$ into Eq.~\xref{f_omega};
the interested reader may consult, \textit{e.g.},
%has been computed elsewhere 
\cite{ulrich-diss10,endlinks-huthm96,gcz96}.

The third term in Eq.~\xref{Ftotal} reflects the coupling of the two order parameters, accounting for phase separation \emph{and} gelation. The lowest order terms allowed by symmetry are explicitly given by
\begin{align}
\lefteqn{ F_{\sigma\Omega}[\sigma,\Omega] = } \label{f_coupled}\\
&=-\frac{(L\chi_{\mu})^2\mu}{8\varrho_0^2}
\sum_{\bm k,\bm p\neq 0} \sigma_{\bm k}\sigma_{\bm p}\Omega({-\bm k,-\bm p}) 
\;s_{\sigma^2\Omega}(k^2,p^2)
\notag\\
&-\frac{L\chi_{\mu}\mu^2}{4\varrho_0}\!\!\sum_{\bm k,\bm p,\bm q\neq 0}\!\!
\sigma_{\bm q}\Omega({\bm k,\bm p}) \Omega({-\bm k-\bm q,-\bm p})
\;s_{\sigma\Omega^2}(\bm q; (\bm k, \bm p) ).
\notag
\end{align}
In a phenomenological approach, the correlations $s_{\sigma^2\Omega}$
and $s_{\sigma\Omega^2}$ are constants, whereas in the microscopic
approach these correlations can be computed from first principles, as
shown in the appendix. In the following sections, we are going to
analyze the free energy and discuss the lamellar state as well as the
instability towards a lamellar gel.

%\subsection{\label{sec:inst_mix_fluid}Instabilities of the mixed fluid}

\section{\label{sec:lam-fluid}Lamellar fluid}
As discussed above, the chemically homogeneous melt becomes unstable to
microphase separation at $ 2 - L\chi_{\mu_c} s(q_0^2)=0$. 
In the vicinity of the transition line, 
the amplitude of the composition modulation
is expected to be small and hence 
appropriately described by a Landau-Wilson expansion of the free-energy functional. 
Assuming a lamellar state [Eq.~\xref{orderparlam}], the free energy becomes
\begin{align}
\lefteqn{
F_{\text {lam}} [\sigma,\Omega]  =  }\\
& \frac{L\chi_{\mu}}{4\varrho_0^2}    
\left( 1 - L\chi_{\mu} \frac{s(q_0^2)}{2} \right) 
\sigma_{0}^2+c_1 \sigma_0^4\notag\\
&+\frac{\mu}{2}\sum_{\bm k,\bm p\neq 0}
\big(1 - 4\mu d_{1/2}(k^2+p^2)\big)
\Omega({\bm k},{\bm p})\Omega({-\bm k},{-\bm p})\notag\\
&-2c_3\sigma_0^2(\Omega({\bm q_0},{\bm q_0})+\Omega({\bm q_0},{-\bm q_0}))
+{\cal O}(\Omega^3,\sigma\Omega^2)
\notag\end{align}
Here $c_1$ and $c_3$ abbreviate the vertex functions $s_{\sigma^4}$ and
$s_{\sigma^2\Omega}$, evaluated
at the lamellar wavevectors $\pm \bm q_0$.

Variation of $F_{\text {lam}}$ with respect to the order-parameter fields
yields saddle-point equations, 
%\textit{inter alia\/} one that results in %predicts prescribes 
which \textit{inter alia\/} show that
the gel order parameter $\Omega$ assumes a nontrivial value
as a function of the amplitude $\sigma_0$ even in the fluid state.
%Also the functional 
The complete theory, %derived from the microscopic model, 
including all moments of $\Omega$, 
cf.\ Appendix~\ref{app:free-energy}, shows that
to lowest order in the amplitude 
$\sigma_0$, indeed only the second moment is affected, 
as announced before Eq.~\xref{f_omega}.
For a lamellar microphase, the saddle-point equation, see also Eq.~\xref{sp-omega}, 
reads
\be \label{lamomega}
\Omega_{\text {lam}}({\bm k},{\bm p})=
\frac{c_3\sigma_0^2(\delta_{\bm k,\bm q_0} + \delta_{\bm k,-\bm q_0}) (\delta_{\bm p,\bm
  q_0} + \delta_{\bm p,-\bm q_0})}{\mu\big(1 - 4\mu
  d_{1/2}(k^2+p^2)\big)},
\ee
valid for $\mu < 1$. 
This implies a periodic modulation of the second moment of the local density of 
$A$ segments, as anticipated in Eq.~\xref{orderparlam}.
Not written out explicitly here, 
generalized compression modes $\kappa\varrho_{\bm k}$ of the simple density are on the saddle-point level also quadratic in $\sigma_{\bm k}$ in the symmetric case.
However, the dominant wave number of this modulation is $2q_0$, which can be explained with the system's tendency to compress both $A$- and $B$-rich regions.

%Variation with respect to $\sigma_0$ and $\Omega$ yields
%\begin{align}
%0 &= \frac{\partial F_{\text {lam}}}{\partial \sigma_0}
%\label{spsigma}\\
%&=\frac{L\chi_{\mu}}{2\varrho_0^2}    
%\left( 1 - L\chi_{\mu} \frac{s(q_0^2)}{2} \right) 
%\sigma_{0}+4c_1\sigma_0^3\notag\\
%&\quad \mbox{} + 4c_3\sigma_0(\Omega({\bm q_0},{\bm q_0})+\Omega({\bm q_0},{-\bm q_0}))\notag\\
%0 &= \frac{\delta F_{\text {lam}}}{\delta \Omega({-\bm k},{-\bm p})}\label{spomega}\\
%&=
%\mu\big(1-\mu d(0.5, k^2+p^2)\big)\Omega({\bm k},{\bm p}) \notag\\
%&\quad\mbox{}-c_3\sigma_0^2
%(\delta_{\bm k,\bm q_0}+\delta_{\bm k,-\bm q_0})
%(\delta_{\bm p,\bm q_0}+\delta_{\bm p,-\bm q_0})
%\notag
%\end{align}
%As long as $\mu<1$, we can solve Eq.~\xref{spomega} for $\Omega({\bm
%  k},{\bm p})$ 
%\be \label{lamomega}
%\Omega_{\text {lam}}({\bm k},{\bm p})=(\delta_{\bm k,\bm
%  q_0}+\delta_{\bm k,-\bm q_0}) (\delta_{\bm p,\bm
%  q_0}+\delta_{\bm p,-\bm q_0})\frac{c_3\sigma_0^2} {\mu\big(1 - 4\mu
%  d(0.5, k^2+p^2)\big)} 
%\ee
%and find a periodic modulation of the second moment of the local density of 
%$A$-monomers. 

Substituting these results %and the generalized compression modes 
back into the equation for the lamellar amplitude $\sigma_0$
%Eq.~\xref{spsigma} 
and solving for $\sigma_0\neq 0$, we find
\be\label{sig0}
\sigma_0^2\propto  1 - L\chi_{\mu} \frac{s(q_0^2)}{2}
\ee
close to the transition. 
As expected, the lamellar amplitude grows continuously like a square root 
as a function of the distance to the critical point
 --- characteristic of a mean-field theory.
For microphase separation 
without coupling to another transition, 
% alone 
 this result was first derived by Leibler~\cite{leibler80}. 
Later, in Ref.~\cite{fred-helfand87}, it was shown that fluctuations induce a first-order transition in the same universality class as the Brazovski\u\i\
model \cite{brazov75,braz-dzyal87}.

Interestingly, the contribution of the second moment of $\Omega$ causes an increase in the amplitude compared to a diblock melt without cross-links.
Due to the wave-number dependence of the denominator of the expression for $\sigma_0$
(omitted in Eq.~\xref{sig0}),
also the lamellar wave number %that maximizes $\sigma_0$ 
generally changes continuously from $q_0$ when the amplitude increases.

We summarize the main results obtained so far: 
In the fluid state,
a phase transition to a lamellar state occurs at a critical value of the
incompatibility. 
Cross-linking causes a pre-ordering which facilitates phase separation, 
so that the critical incompatibility depends on the degree of cross-linking.
Microphase separation and cross-linking are coupled already in
 this state which has not yet undergone gelation. 
 Furthermore, within the
lamellar phase, the gel order parameter is nonzero, even though the
monomers are not strictly localised, but follow a periodic probability
density.

\section{\label{sec:lam-gel}Lamellar gel}
What happens, if we increase the number of cross-links in the lamellar
fluid? At which cross-link density does gelation set in,
what is the structure of the resulting gel, 
and what remains of the microstructure? 
To answer these questions,
we test the stability towards gelation by
inserting small deviations $\delta\Omega$ of the gel order parameter 
from its value in the lamellar fluid,
Eq.~\xref{lamomega}, \textit{viz.},
\be
\Omega({\bm k_1},{\bm k_2})=\Omega_{\text {lam}}({\bm k_1},{\bm k_2})+\delta
\Omega({\bm k_1},{\bm k_2}),
\ee
into the free-energy functional and expanding the free-energy difference 
\be
\Delta F = F[\sigma,\Omega]
- F_{\text {lam}}[\sigma,\Omega]
\ee
up to quadratic order in $\delta\Omega$.
In doing this, we insert the form of the novel component of the gel order parameter
dictated by the discrete translation symmetry of the lamellar composition pattern, \textit{cf.}, Eq.~\xref{orderparlam-gel}.

%\subsection{\label{Matrix2} Quadratic form indexed by gel components
%}
%An alternative method consists in 
As suggested by the form of the contributions to the gel order parameter discussed in 
Sec.~\ref{sec:orderpar},
we decompose the %gel 
deviation (vector) $\bm\delta\Omega$
into %its isotropic, homogeneous and its lamellar parts, viz.,
orthogonal components that represent isotropic,
homogeneous, respectively sinusoidal density of localised $A$ segments:
\begin{align}
\lefteqn{\bm{\delta} \Omega (\bm k_1, \bm k_2) = }
\\
& \delta\Omega^{(\text i)} \delta_{\bm k_1 + \bm k_2, \bm 0}
+ \delta\Omega^{(+)} \delta_{\bm k_1 +  \bm k_2, +\bm q_0}
+ \delta\Omega^{(-)} \delta_{\bm k_1 +  \bm k_2, -\bm q_0}.
\notag
\end{align}
Since the lamellar modulation affects one space dimension (defined by $\bm q_0$, labeled 1) only, 
we focus on the component $k \mathrel{\mathop:}= k_{1,1}$ %, $p_{1,1}$ 
of the variable wavevector $\bm k_1$. %and $\bm p_1$
%As written out explicitly in Eq.~\xref{Matrix1}, 
Additionally, only one of the wave numbers $k_{1,1}$ and $p_{1,1}$ 
%are again constrained by 
is a free parameter, called $k$,
due to translational invariance of the conformational average, \textit{cf.}, Eq.~\xref{f_coupled}.
The resulting quadratic form
\be
\Delta F =
\sum_{m,n \in \{ \text i, +, - \} } G^{(m,n)}(k) \delta\Omega^{(m)} \delta\Omega^{(n)},
\ee
%\be
%G^{(\text i, \text i)}(k) =
%\ee
parametrized by $k$ can be further simplified by taking the limit $k\to 0$, 
anticipating the gel instability to set in with a small critical wave number
(without lamellar order, the gel instability occurs at wave number $k_{\text{gel}}=0$).
Coupling of these matrices for each $k$ and thus 
a more complicated diagonalisation would arise 
but for $k \geq 2 q_0$.
Still, the description includes the lamellar wave number $q_0$, which can be 
reasonably assumed to be much larger than the gel instability wave number.
In this limit, the $(+)$ and $(-)$ wave contributions collapse to a single one, and the problem is %approximately 
reduced to computing the eigenvalues of the 
symmetric $2\times2$ instability matrix
\begin{align}
\label{matrix-gel}
%\lefteqn{ 
G = %}\\
%%\mu\times
%&
\left( \begin{array}{cc}
1 - \mu 
& -\mu_{\sigma_0} s_{\sigma\Omega^2}(q_0^2)
%-\frac{\displaystyle \mu L\chi_{\mu} \sigma_0 s_{\sigma\Omega^2}(q_0^2) }{\displaystyle2\varrho_0} 
\\
-\mu_{\sigma_0} s_{\sigma\Omega^2}(q_0^2)
 & 2 \left( 1 - 4\mu d_{1/2}(q_0^2) \right)
\end{array}
\right),
%\notag
\end{align}
with the amplitude-dependent quantity %rescaled cross-link concentration
\be
\mu_{\sigma_0}\mathrel{\mathop:}= \frac{\mu L\chi_{\mu}\sigma_0}{2\varrho_0}
\ee
The mixed vertex $s_{\sigma\Omega^2}$, Eq.~\xref{s_sigOm2}, evaluated at $k=0$ is
\begin{align}
s_{\sigma\Omega^2}(q_0^2) & = 2\int_0^1\!\d s_1 q(s_1) \int_0^{1/2}\!\d s_2 \e^{|s_2 - s_1|q_0^2} \\
& = 4 d_{1/2}(q_0^2) - d_{1}(q_0^2),
\notag
\end{align}
\textit{i.e.}, identical to the quadratic correlation function of the unconstrained diblock melt Eq.~\xref{s_symm}. 

In order to determine the stability towards gelation, we have to consider
the smaller eigenvalue of $G$, Eq.~\xref{matrix-gel},
%indicates the instability toward the lamellar gel is
\begin{align}
2\lambda_{c} = &\, 3 - \mu (1 + 8 d_{1/2}(q_0^2) )
\\
& \mbox{} - \sqrt{
\left[1 + \mu(1- 8 d_{1/2}(q_0^2) )\right]^2 + \left[2 \mu_{\sigma_0}
%\mu L\chi_{\mu}\frac{\sigma_0}{\varrho_0}  
s_{\sigma\Omega^2}(q_0^2)\right]^2
},
\notag\end{align}
and spot its change of sign. 
By consistently computing  the lamellar amplitude $\sigma_0$ and wave number $q_0$ for each set $(\mu, L\chi_{\mu})$ 
and the corresponding value of $\lambda_c$ and varying $L\chi$ until the latter approaches zero,
we trace the lamellar gel instability $\mu_c (L\chi)$ 
shown as the dashed line in in Fig.~\ref{fig:phased}.
The end point toward small $\mu$ in this graph arises from % is determined by the
the limitation of the approach to
%As visible there, the approach is limited to 
the vicinity of the fluid lamellar instability by the assumption that the amplitude $\sigma_0$ be small.

To summarize, the impact of microphase separation in the fluid
%combined with selective cross-linking 
is to shift the gelation transition to smaller critical cross-link chemical potentials.
Microphase separation creates domains enriched in the selected component, 
%selected for cross-linking, 
so that 
cross-linking %in one type of domains
in the presence of $AB$ ordering is more efficient than without. %$AB$ ordering.
This effect can be captured due to the cross-link ensemble, Eq.~\xref{DE-distrib}, which %weighs network realizations according to spatial correlations.
assigns a higher probability to network realizations with many node pairs close to each other in space at the instant of cross-linking.

\section{\label{sec:conclusion}Conclusions and Outlook}

In this work, we have mapped out %studied 
the %multiple 
phase states and transition lines in an $AB$ diblock copolymer melt cross-linked at random 
by irreversible and type-selective bonds.
The central questions underlying this %theoretical 
study %we aimed to clarify 
were:
How does random, type-selective cross-linking affect the chemical ordering transition? 
Vice versa, how does the ordering affect the gelation transition?
Is there a novel transitions from the ordered fluid to the ordered gel?
%Explore %coupling 
How do the different order parameters interact in a minimal field-theoretical description?

The resulting phase diagram is rather complex due to the symmetries of
the underlying Hamiltonian. 
First, translational invariance is spontaneously broken in the gel state 
due to random localisation of particles. 
%At a first glance,
Secondly, considered superficially, microphase separation of symmetric diblocks
is in the Ising universality class, and mean-field theory
predicts a second-order phase transition. 
However, a more careful
analysis of fluctuations has shown that microphase separation is
instead in the Brazovski\u\i\ universality class for weak crystallization 
\cite{brazov75,braz-dzyal87} and exhibits a
fluctuation-induced first-order transition \cite{fred-helfand87}. 
Since our focus here has been on the \textit{coupling} of microphase separation and gelation, we have restricted ourselves to the mean-field theory.
Selective cross-linking gives rise to a partial pre-ordering into $A$- and $B$-rich domains, 
resulting in a smaller critical $AB$ incompatibility
$\chi_c$ for microphase separation. 
Still, the Brazovski\u\i\
mechanism can be expected to be at work due to the shell of nonzero wavevectors which characterise the instability toward microphase separation also at these altered  
$\chi_c$.
Further studies beyond mean field are needed to confirm this conjecture. 
Microphase separation, here into lamellae, facilitates the formation of a gel, because 
the success of cross-linking $A$-selectively is enhanced by the existence of $A$-rich domains (the same argument holds for $B$-selective cross-links). Therefore,
fewer cross-links are needed to stabilise the network.

Several extensions of our work can and should be pursued. So far we
have only considered isotropic localisation, which might be
approximately valid for very weak segregation. However in general,
localisation will be anisotropic, with strong localisation within the
lamellar domains enriched in $A$ and weak localisation perpendicular to the
lamellar plane. Our ansatz can be easily generalised to account for
such anisotropies with different localisation lengths for localisation
in plane and in between planes.

Based on the equilibrium analysis of this work, many striking features
can be expected when turning to the nonequilibrium behavior. 
One especially interesting route to nonequilibrium is offered by the
possibilty to cross-link
the system at one temperature ($1/\chi_p$) and then measure its
properties at a different temperature ($1/\chi_m$). 
For example, one can cross-link the sample within the lamellar phase, then decrease
$\chi$ and study to what extent the system is able to (re)mix 
--- in dependence on the cross-link density. 
Experimentally, it is possible to open cross-links by UV light,
so that the system is allowed to partially relax, depending
on the amount of opened cross-links. 
In general, phase ordering will be frustrated due to constraints imposed 
onto the system at a different degree of ordering. 
%Or does 
Yet, is it possible that a structure of frustrated order emerges
%Which structure arises 
when $\chi$ is increased in the disordered gel?
Theoretically one can again use the
Deam-Edwards distribution to allow for the analytical treatment of
different states at preparation and measurement. 
Such studies have already been performed for cross-linked polymer
blends~\cite{wald-zipp-goldb-epl05}, and work is in progress for the
system considered here.

Another promising direction is to analyze the elastic properties
of gel states with periodical composition modulation, 
in this study exemplified by lamellar gels.
These states %are reminiscent of smectic liquid crystals with 
are expected to display special mechanical properties 
analogous to smectic liquid crystals,
since the gel component is amorphous in all but one spatial direction,
where the periodic composition modulation occurs. %in one spatial direction.
The difference is that the centres of mass in the gel component are  
%not mobile as in a liquid, but 
randomly trapped %localised 
such as in a glass. 
Topological constraints due to cross-linking may produce more complicated phase states or even phase coexistence in a block copolymer melt
with sequence heterogeneity \cite{hmz11PRE}.

Finally, as alluded to above, 
it would be desirable to go beyond mean-field theory,
also to study coupled fluctuations of the two order
parameters for localisation and for microphase separation. 
This is particularly interesting due to the drastic
effects which fluctuations have already on microphase separation alone.

\begin{acknowledgments}
Financial support of this work by the Deutsche Forschungsgemeinschaft through grant SFB-937/A4 is gratefully acknowledged. 
We thank K.~Tietz, P.~Vana, S.~Finkh\"auser, and K.~Samwer from project A4 for the fruitful collaboration and M.~M\"uller for interesting discussions.
\end{acknowledgments}

\appendix
\section{\label{app:free-energy}
Route to the free-energy functional from the microscopic description
%Effective Hamiltonian for incompressibility and rigid cross-links
}

Here, we sketch the calculation of the %complete
effective Hamiltonian or free-energy functional based on the microscopic model,
which \textit{inter alia} provides us with the explicit form of the vertices. 
First, we decouple the pair interactions quadratic in the collective densities, cf.~Eqs.~\xref{H-ev} and \xref{H-chi}, 
with (auxiliary) interaction fields $\tilde\sigma$, $\tilde \varrho$, and $\tilde \Omega$ using Hubbard-Stratonovich transforms. 
Then, we perform the average $\left[\cdot \right]_{\mathcal{P}_{\text{DE}}}$ 
with the Deam-Edwards cross-link distribution Eq.~\xref{DE-distrib} 
by means of the replica method \cite{mezard}
and obtain the replica partition function $\tilde{\mathcal{Z}}_{n+1}$. 
The latter involves $n+1$ replicas or system copies, since
$\mathcal{P}_{\text{DE}}$ requires an extra replica at %of the 
preparation %state
(labeled $\alpha=0$ in the following), 
%characteristic of  the Deam-Edwards distribution,
%cf.\ the discussion after Eq.~\xref{DE-distrib},  
in addition to the $n$ replicas at %of the 
measurement %state 
(labeled $\alpha=1,\ldots, n$) 
usually employed to reformulate the average of $\ln {\cal Z}$.
From $\tilde{\mathcal{Z}}_{n+1}$, 
the disorder-averaged free energy $\left[F\right]_{\mathcal{P}_{\text{DE}}}$
follows as 
\be 
- \left[F\right]_{\mathcal{P}_{\text{DE}}} =
\lim_{n\to 0} \frac{\tilde{\mathcal{Z}}_{n+1} - \tilde{\mathcal{Z}_{1}}}{n \tilde{\mathcal{Z}_{1}}}
\ee
where $\lim_{n\to 0} \tilde{\mathcal{Z}}_{n+1} = \tilde{\mathcal{Z}_{1}}$.
\begin{subequations}
\begin{align}
\lefteqn{ \tilde{\mathcal{Z}}_{n+1}  = }\label{replica-Z}\\ 
&\int\!{\cal D}\left[\tilde\sigma^{(\alpha)},\tilde\varrho^{(\alpha)}\right]\int\!{\cal D}\left[\tilde\Omega \right]
 \exp\left\{ - N \tilde h_{n+1} \left[\tilde\sigma^{(\alpha)},\tilde\varrho^{(\alpha)},\tilde \Omega\right] \right\}.
 \notag
 \end{align}
Herein, $\tilde\sigma$ and $\tilde\varrho$ are, respectively, the interaction fields for the imbalance and the total density, confined to the 1 replica sector (1RS), 
and $\tilde\Omega$ the field interacting with the gel order parameter, which according to Eq.~\xref{OP2replica} has $\geq 2$ arguments $\bm k, \bm p,\dots$ and thus is confined to the higher replica sector (HRS).
The effective Hamiltonian which governs these field reads
\begin{align}
\lefteqn{ 
 \tilde h_{n+1}  = }\label{h_eff_n+1}\\
 &\frac{L}{4\varrho_0^2} \sum_{\alpha=0}^n   \sum_{\bm k\neq \bm 0} 
\left[
\chi^{(\alpha)}_{\mu} \tilde\sigma_{\bm k}^{(\alpha)} \tilde\sigma_{-\bm k}^{(\alpha)}
+ \kappa^{(\alpha)}  \tilde\varrho_{\bm k}^{(\alpha)} \tilde\varrho_{-\bm k}^{(\alpha)}
\right]
\notag\\
&\mbox{} + \frac{\mu}{2} \sum_{\hat k\in\text{HRS}} \tilde\Omega_{\hat k} \tilde\Omega_{-\hat k}
- \ln \tilde z_{n+1} \left[\tilde\sigma^{(\alpha)},\tilde\varrho^{(\alpha)},\tilde \Omega\right] 
\notag
 \end{align}
with the single-polymer partition function
\begin{widetext}
\begin{align}
\lefteqn{ \tilde z_{n+1} \left[\tilde\sigma^{(\alpha)},\tilde\varrho^{(\alpha)},\tilde \Omega\right] = }\\
&\Biggl\langle \exp\left\{
\frac{1}{2\varrho_0} \sum_{\alpha=0}^n 
\smash{ \sum_{\bm k\neq \bm 0} \int_0^1\!\d s }
\left[
L\chi^{(\alpha)}_{\mu} \tilde\sigma_{\bm k}^{(\alpha)} \left[ q(s) -\mathvar{q}\right]
+ i L\kappa^{(\alpha)}  \tilde\varrho_{\bm k}^{(\alpha)}
\right] \e^{-i\bm k\cdot\bm r^{(\alpha)}(s)} 
%\right.\Biggr.
%\notag\\
%& \quad\quad\quad\quad \quad \Biggl.\left.  \mbox{} 
+ 2 \mu \smash{ \sum_{\hat k\in\text{HRS}} } \tilde\Omega_{\hat k}  \int_0^{1/2} \!\d s \,
\e^{-i \hat k \cdot \hat r(s) }
\right\}
\Biggr\rangle_{\!\hat{\mathcal{H}}_{\text{G}} }
\notag
\end{align}
\end{widetext}
\end{subequations}
Hatted wave- and position vectors are used as a shorthand for arrays of $n+1$ vectors in replica space, $\hat k \mathrel{\mathop:} = (\bm k^{(0)}, \bm k^{(1)},\ldots, \bm k^{(n)})$, and $\hat k\cdot \hat r = \sum_{\alpha = 0}^n \bm k^{(\alpha)} \cdot \bm r^{(\alpha)}$. 
Accordingly, here and in the following, $\langle\cdot\rangle_{\!\hat{\mathcal{H}}_{\text{G}} }$ denotes the $n+1$-fold replicated single-chain conformational average. 
For the sake of transparency, we present the computation for $AB$-exchange symmetric melts, rigid cross-links ($a_c\to 0$), and incompressibility.
In this case, a simple relation holds between the expectation values of the collective densities introduced in Sec.~\ref{subsec:diblocks} and of the interaction fields:
\begin{align}
\left\langle \sigma_{\bm k} %^{(\alpha)} 
\right\rangle_{\mathcal{H}} & = \lim_{n\to 0} \left\langle \tilde\sigma_{\bm k}^{(\alpha)} \right\rangle_{N\tilde h_{n+1}}, \notag\\
\left\langle \varrho_{\bm k}%^{(\alpha)} 
\right\rangle_{\mathcal{H}} &= i  \lim_{n\to 0} \left\langle \tilde\varrho_{\bm k}^{(\alpha)} \right\rangle_{N\tilde h_{n+1}}
\end{align}
(analogously for $\tilde \Omega$), so that on the saddle-point level we can identify these fields with the order parameters.
Therefore, we return to the notation $\sigma$, $\varrho$, $\Omega$ for the fields.
One further simplification in this case is that the elimination of total density fluctuations at the saddle point
%in the 1RS
does not affect the instability towards phase separation.

%\section{Saddle-point equations}

\section{\label{app:vertices}Vertices and saddle-point equations}

The complete set $s_{\sigma^4}$ of fourth-order vertices in $\sigma$ reads
\begin{widetext}
\begin{align}
\sideset{}{'}\sum_{\bm k_1, \bm k_2, \bm k_3
}\!\!s_{\sigma^4} %(\bm k_1, \bm k_2, \bm k_3)
=&  \frac{L^4}{3\cdot 2^7\varrho_0^4} \sum_{\alpha=0}^n \left( \chi^{(\alpha)}_{\mu} \right)^4
\left\{
\sum_{\bm k_1, \bm k_2 \neq \bm 0} 
\Bigl[ 3 s(k_1^2) s(k_2^2) - s_{4} \left( \bm k_1, - \bm k_1, \bm k_2 \right) \Bigr]
\sigma_{\bm k_1}^{(\alpha)}\sigma_{-\bm k_1}^{(\alpha)}
\sigma_{\bm k_2}^{(\alpha)}\sigma_{-\bm k_2}^{(\alpha)}
\right.
\notag\\
& 
\left. \mbox{} + 
\!\!\!\sideset{}{'}\sum_{\bm k_1 \neq -\bm k_2, \bm k_3
%\begin{array}{l} \scriptstyle \bm k_1, \bm k_2, \bm k_3 \\ \scriptstyle \bm k_1+\bm k_2 \neq \bm 0 \end{array}
}\!
\biggl[
3 \frac{ s_{\sigma^2\varrho}(\bm k_1, \bm k_2) 
s_{\sigma^2\varrho}(\bm k_3, -\sum_{\text{r}=1}^3\bm k_{\text r}) 
}{d_1\left(\left(\bm k_1 + \bm k_2\right)^2 \right)}
- s_{4} \left( \bm k_1, \bm k_2, \bm k_3 \right)
\biggl]
\sigma_{\bm k_1}^{(\alpha)}\sigma_{\bm k_2}^{(\alpha)}
\sigma_{\bm k_3}^{(\alpha)}\sigma_{-\sum_{\text{r}=1}^3\bm k_{\text r}}^{(\alpha)}
\right\}
\end{align}
\end{widetext}
with the correlation function $s_{\sigma^2\varrho}$ 
%from the elimination of non-critical 
due to the contribution of generalized density modulations,
\begin{align}\label{s_rhosigma2}
\lefteqn{s_{\sigma^2\varrho}\left(\bm k_1, \bm k_2 \right) }\\
&\mathrel{\mathop:}= \iiint_{0}^1 \!\!\d s_1 \d s_2 \d s_3\,
q(s_1)q(s_2) \notag\\
&\quad\times
\left\langle\e^{-i \left(\vphantom{\int} \bm k_1 \cdot \bm r^{(\alpha)}(s_1) + \bm k_2 \cdot \bm r^{(\alpha)}(s_2) + \bm k_3 \cdot \bm r^{(\alpha)}(s_3)
    \right)}\right\rangle_{\!\hat{\mathcal{H}}_{\text{G}} }\notag\\
& = \iiint_{0}^1 \!\!\d s_1 \d s_2 \d s_3 q(s_1)q(s_2)\notag\\
&\quad\times
\,\e^{ 
 |s_2-s_1|\bm k_1\cdot\bm k_2 
- |s_3-s_2|\left(\bm k_1\cdot\bm k_2 + k_2^2\right) 
- |s_3-s_1|\left(\bm k_1\cdot\bm k_2 + k_1^2\right)
}
\notag\end{align}
(this function has been computed as $s^{(\alpha)}(\bm k_1, \bm k_2)$ in \cite{alicePhD}, also for diblocks).
The explicit form of the fourth-order correlation $s_4(\bm k_1, \bm k_2, \bm k_3)$
evaluated for lamellar phase separation and for symmetric diblocks
can be found in \cite{alicePhD}, 
denoted as $s^{(\beta)}$. 
%for the set $(\bm k, \bm k, -\bm k)$ relevant to the lamellar state

The general saddle-point equation which determines the second moment of the gel order parameter as a function of the order parameter for microphase separation is
\begin{align}
\overline{\Omega_{(\bm k_1,\bm k_2)}} =
%\mbox{}
\frac{ \displaystyle (L\chi_{\mu})^2 
s_{\sigma^2\Omega}(k_1^2, k_2^2)
\overline{\sigma_{\bm k_1}}\, \overline{\sigma_{\bm k_2}} }{\displaystyle
8\varrho_0^2\left(1 - \mu \vphantom{\int} \overline{d_{1/2}(k_1^2+ k_2^2)}\right)} 
+ {\cal{O}} (|\sigma|^3), 
\label{sp-omega}
\end{align}
where we assumed replica-symmetry.
The correlation function $s_{\sigma^2\Omega}(k_1^2, k_2^2)$ in Eq.~\xref{sp-omega},
first appearing in Eq.~\xref{f_coupled}, 
for $\hat k = \bm k_1 \hat e^{(\alpha)} + \bm k_1 \hat e^{(\beta)}  $ and $\alpha\neq\beta$
is
\begin{align}\label{s_Omegasigma2}
\lefteqn{
s_{\sigma^2\Omega}\left(k_1^2, k_2^2 \right) }\\
& \mathrel{\mathop:}= 2 \iint_{0}^1 \!\d s_1 \d s_2 q(s_1)q(s_2)\int_{s_3 = 0}^{1/2} \! \d s_3
\notag\\
&\quad \times \left\langle
\e^{-i \left(\vphantom{\int} \bm k_1 \cdot \bm r^{(\alpha)}(s_1) + \bm k_2 \cdot \bm r^{(\beta)}(s_2) + \hat k \cdot \hat r(s_3)\right)}
\right\rangle_{\!\hat{\mathcal{H}}_{\text{G}} }\notag\\
& =  2\iint_{0}^1\!\d s_1 \d s_2 q(s_1)q(s_2) \int_{s_3 = 0}^{1/2} \! \d s_3
\notag\\
&\quad \times
 \e^{ - |s_1-s_3|k_1^2 - |s_2-s_3|k_2^2} %\quad(\alpha\neq\beta)
\notag.
\end{align}

With the contributions of the generalized density modulations and the gel order parameter, the fluid lamellar amplitude is determined by
\begin{align}
\lefteqn{\left( \frac{L\chi_{\mu} \sigma_0}{4 \varrho_0} \right)^2  = \Phi(q_m) 
= \max_{k} \Phi(k),
} \\
%\lefteqn{ 
&\Phi(k) \mathrel{\mathop:}=
%} \notag\\& 
\frac{ \displaystyle \frac{s(k^2)}{2} - \frac{1}{L\chi_{\mu}} }{%
D(k^2)
%2 \left( s(k^2) \right)^2 + \displaystyle{\frac{ \left( s_{\sigma^2\varrho} (k\bm n, k\bm n) \right)^2 }{d_1(4k^2)} - s_{4} \left( k\bm n , k \bm n, - k \bm n \right) - 2\mu  \frac{\left( s_{\sigma^2\Omega}(k^2, k^2) \right)^2}{1 - 4 \mu \vphantom{\int} d_{1/2}(k^2 + k^2)} }
},
\notag\\
& L\chi_{\mu} \geq L\chi_{\mu, c} = \frac{2}{s(q_0^2)}, \quad \mu < 1,
 \quad q_{\text m} \mathrel{\mathop:}= \underset{k}{\text{argmax}}\, \Phi(k),
\notag
\end{align}
with the denominator
\begin{align}
D(k^2) = &\, 2 \left( s(k^2) \right)^2 
+ \frac{ \left( s_{\sigma^2\varrho} (k\bm n, k\bm n) \right)^2 }{d_1(4k^2)} \\
& \mbox{} - s_{4} \left( k\bm n , k \bm n, - k \bm n \right) 
- 2\mu  \frac{\left( s_{\sigma^2\Omega}(k^2, k^2) \right)^2}{1 - 4 \mu d_{1/2}(2 k^2)},
\notag
\end{align}
and $\bm n$ the arbitrary, but fixed unit normal defined by $\bm q_0$.

The other correlation function in Eq.~\xref{f_coupled} is
\begin{align}
\lefteqn{ s_{\sigma\Omega^2} (\bm k^{(\alpha)}; \hat k)  = }\label{s_sigOm2}\\ 
& 4\int_0^1\d s_1 q(s_1) \iint_0^{1/2}\!\!\d s_2\d s_3\notag\\
&\quad\times
\e^{
|s_2 - s_1|\bm k^{(\alpha)}\cdot\hat k
-|s_3-s_1|\left( \bm k^{(\alpha)}\cdot\hat k + k^2 \right)
-|s_3-s_2|\left( \bm k^{(\alpha)}\cdot\hat k + \hat k^2 \right)
},
\notag
\end{align}
and is essential %crucial 
to determine the instability toward the lamellar gel in Sec.~\ref{sec:lam-gel}.

%The correlation function $c_1$ is given by
The third-order correlation for the gel order parameter in Eq.~\xref{f_omega} is
\begin{align}
s_{\Omega^3}(\hat k_1, \hat k_2) = &
8\iiint_{0}^{1/2} \! \d s_1 \d s_2\d s_3\,\e^{
|s_2-s_1| \hat k_1 \cdot \hat k_2 }\\
&\times\e^{
- |s_3-s_1|\left( \hat k_1 \cdot \hat k_2 + \hat k_1^2 \right)
- |s_3-s_2|\left( \hat k_1 \cdot \hat k_2 + \hat k_2^2 \right)
}.
\notag\end{align}

\bibliography{XlinkedDiblocks}

%merlin.mbs aipnum4-1.bst 2010-07-25 4.21a (PWD, AO, DPC) hacked
%Control: key (0)
%Control: author (8) initials jnrlst
%Control: editor formatted (1) identically to author
%Control: production of article title (-1) disabled
%Control: page (0) single
%Control: year (1) truncated
%Control: production of eprint (0) enabled
\begin{thebibliography}{26}%
\makeatletter
\providecommand \@ifxundefined [1]{%
 \@ifx{#1\undefined}
}%
\providecommand \@ifnum [1]{%
 \ifnum #1\expandafter \@firstoftwo
 \else \expandafter \@secondoftwo
 \fi
}%
\providecommand \@ifx [1]{%
 \ifx #1\expandafter \@firstoftwo
 \else \expandafter \@secondoftwo
 \fi
}%
\providecommand \natexlab [1]{#1}%
\providecommand \enquote  [1]{``#1''}%
\providecommand \bibnamefont  [1]{#1}%
\providecommand \bibfnamefont [1]{#1}%
\providecommand \citenamefont [1]{#1}%
\providecommand \href@noop [0]{\@secondoftwo}%
\providecommand \href [0]{\begingroup \@sanitize@url \@href}%
\providecommand \@href[1]{\@@startlink{#1}\@@href}%
\providecommand \@@href[1]{\endgroup#1\@@endlink}%
\providecommand \@sanitize@url [0]{\catcode `\\12\catcode `\$12\catcode
  `\&12\catcode `\#12\catcode `\^12\catcode `\_12\catcode `\%12\relax}%
\providecommand \@@startlink[1]{}%
\providecommand \@@endlink[0]{}%
\providecommand \url  [0]{\begingroup\@sanitize@url \@url }%
\providecommand \@url [1]{\endgroup\@href {#1}{\urlprefix }}%
\providecommand \urlprefix  [0]{URL }%
\providecommand \Eprint [0]{\href }%
\providecommand \doibase [0]{http://dx.doi.org/}%
\providecommand \selectlanguage [0]{\@gobble}%
\providecommand \bibinfo  [0]{\@secondoftwo}%
\providecommand \bibfield  [0]{\@secondoftwo}%
\providecommand \translation [1]{[#1]}%
\providecommand \BibitemOpen [0]{}%
\providecommand \bibitemStop [0]{}%
\providecommand \bibitemNoStop [0]{.\EOS\space}%
\providecommand \EOS [0]{\spacefactor3000\relax}%
\providecommand \BibitemShut  [1]{\csname bibitem#1\endcsname}%
\let\auto@bib@innerbib\@empty
%</preamble>
\bibitem [{\citenamefont {Leibler}(1980)}]{leibler80}%
  \BibitemOpen
  \bibfield  {author} {\bibinfo {author} {\bibfnamefont {L.}~\bibnamefont
  {Leibler}},\ }\href@noop {} {\bibfield  {journal} {\bibinfo  {journal}
  {Macromolecules}\ }\textbf {\bibinfo {volume} {13}},\ \bibinfo {pages} {1602}
  (\bibinfo {year} {1980})}\BibitemShut {NoStop}%
\bibitem [{\citenamefont {Bates}\ and\ \citenamefont
  {Fredrickson}(1990)}]{bates-fred90rev}%
  \BibitemOpen
  \bibfield  {author} {\bibinfo {author} {\bibfnamefont {F.~S.}\ \bibnamefont
  {Bates}}\ and\ \bibinfo {author} {\bibfnamefont {G.~H.}\ \bibnamefont
  {Fredrickson}},\ }\href@noop {} {\bibfield  {journal} {\bibinfo  {journal}
  {Annu.\ Rev.\ Phys.\ Chem.}\ }\textbf {\bibinfo {volume} {41}},\ \bibinfo
  {pages} {525} (\bibinfo {year} {1990})}\BibitemShut {NoStop}%
\bibitem [{\citenamefont {Matsen}\ and\ \citenamefont
  {Schick}(1994)}]{matsenPRL94}%
  \BibitemOpen
  \bibfield  {author} {\bibinfo {author} {\bibfnamefont {M.~W.}\ \bibnamefont
  {Matsen}}\ and\ \bibinfo {author} {\bibfnamefont {M.}~\bibnamefont
  {Schick}},\ }\href@noop {} {\bibfield  {journal} {\bibinfo  {journal} {Phys.\
  Rev.\ Lett.}\ }\textbf {\bibinfo {volume} {72}},\ \bibinfo {pages} {2660}
  (\bibinfo {year} {1994})}\BibitemShut {NoStop}%
\bibitem [{\citenamefont {Chakraborty}(2001)}]{chakrab2001}%
  \BibitemOpen
  \bibfield  {author} {\bibinfo {author} {\bibfnamefont {A.~K.}\ \bibnamefont
  {Chakraborty}},\ }\href@noop {} {\bibfield  {journal} {\bibinfo  {journal}
  {Phys.~Rep.}\ }\textbf {\bibinfo {volume} {342}},\ \bibinfo {pages} {1}
  (\bibinfo {year} {2001})}\BibitemShut {NoStop}%
\bibitem [{\citenamefont {Sfatos}\ and\ \citenamefont
  {Shakhnovich}(1997)}]{sfatos97rev}%
  \BibitemOpen
  \bibfield  {author} {\bibinfo {author} {\bibfnamefont {C.~D.}\ \bibnamefont
  {Sfatos}}\ and\ \bibinfo {author} {\bibfnamefont {E.~I.}\ \bibnamefont
  {Shakhnovich}},\ }\href@noop {} {\bibfield  {journal} {\bibinfo  {journal}
  {Phys.~Rep.}\ }\textbf {\bibinfo {volume} {288}},\ \bibinfo {pages} {77}
  (\bibinfo {year} {1997})}\BibitemShut {NoStop}%
\bibitem [{\citenamefont {Wald}, \citenamefont {Zippelius},\ and\ \citenamefont
  {Goldbart}(2005)}]{wald-zipp-goldb-epl05}%
  \BibitemOpen
  \bibfield  {author} {\bibinfo {author} {\bibfnamefont {C.}~\bibnamefont
  {Wald}}, \bibinfo {author} {\bibfnamefont {A.}~\bibnamefont {Zippelius}}, \
  and\ \bibinfo {author} {\bibfnamefont {P.~M.}\ \bibnamefont {Goldbart}},\
  }\href@noop {} {\bibfield  {journal} {\bibinfo  {journal} {Europhys.~Lett.}\
  }\textbf {\bibinfo {volume} {70}},\ \bibinfo {pages} {843} (\bibinfo {year}
  {2005})}\BibitemShut {NoStop}%
\bibitem [{\citenamefont {Wald}, \citenamefont {Goldbart},\ and\ \citenamefont
  {Zippelius}(2006)}]{wald-goldb-zipp06}%
  \BibitemOpen
  \bibfield  {author} {\bibinfo {author} {\bibfnamefont {C.}~\bibnamefont
  {Wald}}, \bibinfo {author} {\bibfnamefont {P.~M.}\ \bibnamefont {Goldbart}},
  \ and\ \bibinfo {author} {\bibfnamefont {A.}~\bibnamefont {Zippelius}},\
  }\href@noop {} {\bibfield  {journal} {\bibinfo  {journal} {J.\ Chem.\ Phys.}\
  }\textbf {\bibinfo {volume} {124}},\ \bibinfo {pages} {214905} (\bibinfo
  {year} {2006})}\BibitemShut {NoStop}%
\bibitem [{\citenamefont {Burge}, \citenamefont {Fowler},\ and\ \citenamefont
  {Reaveley}(1977)}]{burge77}%
  \BibitemOpen
  \bibfield  {author} {\bibinfo {author} {\bibfnamefont {R.~E.}\ \bibnamefont
  {Burge}}, \bibinfo {author} {\bibfnamefont {A.~G.}\ \bibnamefont {Fowler}}, \
  and\ \bibinfo {author} {\bibfnamefont {D.~A.}\ \bibnamefont {Reaveley}},\
  }\href@noop {} {\bibfield  {journal} {\bibinfo  {journal} {J.\ Mol.\ Biol.}\
  }\textbf {\bibinfo {volume} {117}},\ \bibinfo {pages} {927} (\bibinfo {year}
  {1977})}\BibitemShut {NoStop}%
\bibitem [{\citenamefont {Rogers}, \citenamefont {Perkins},\ and\ \citenamefont
  {Ward}(1980)}]{rogers80book}%
  \BibitemOpen
  \bibfield  {author} {\bibinfo {author} {\bibfnamefont {H.~J.}\ \bibnamefont
  {Rogers}}, \bibinfo {author} {\bibfnamefont {H.~R.}\ \bibnamefont {Perkins}},
  \ and\ \bibinfo {author} {\bibfnamefont {J.~B.}\ \bibnamefont {Ward}},\
  }\href@noop {} {\emph {\bibinfo {title} {Microbial cell walls and
  membranes}}}\ (\bibinfo  {publisher} {Chapman and Hall},\ \bibinfo {address}
  {London},\ \bibinfo {year} {1980})\BibitemShut {NoStop}%
\bibitem [{\citenamefont {Goldbart}, \citenamefont {Castillo},\ and\
  \citenamefont {Zippelius}(1996)}]{gcz96}%
  \BibitemOpen
  \bibfield  {author} {\bibinfo {author} {\bibfnamefont {P.~M.}\ \bibnamefont
  {Goldbart}}, \bibinfo {author} {\bibfnamefont {H.}~\bibnamefont {Castillo}},
  \ and\ \bibinfo {author} {\bibfnamefont {A.}~\bibnamefont {Zippelius}},\
  }\href@noop {} {\bibfield  {journal} {\bibinfo  {journal} {Adv.\ Phys.}\
  }\textbf {\bibinfo {volume} {45}},\ \bibinfo {pages} {393} (\bibinfo {year}
  {1996})}\BibitemShut {NoStop}%
\bibitem [{\citenamefont {Panyukov}\ and\ \citenamefont
  {Rabin}(1996)}]{panyuk-rabin96}%
  \BibitemOpen
  \bibfield  {author} {\bibinfo {author} {\bibfnamefont {S.~V.}\ \bibnamefont
  {Panyukov}}\ and\ \bibinfo {author} {\bibfnamefont {Y.}~\bibnamefont
  {Rabin}},\ }\href@noop {} {\bibfield  {journal} {\bibinfo  {journal} {Physics
  Reports}\ }\textbf {\bibinfo {volume} {269}},\ \bibinfo {pages} {1} (\bibinfo
  {year} {1996})}\BibitemShut {NoStop}%
\bibitem [{\citenamefont {Tietz}\ \emph {et~al.}(2014)\citenamefont {Tietz},
  \citenamefont {Finkh\"auser}, \citenamefont {Samwer},\ and\ \citenamefont
  {Vana}}]{Tietz14}%
  \BibitemOpen
  \bibfield  {author} {\bibinfo {author} {\bibfnamefont {K.}~\bibnamefont
  {Tietz}}, \bibinfo {author} {\bibfnamefont {S.}~\bibnamefont {Finkh\"auser}},
  \bibinfo {author} {\bibfnamefont {K.}~\bibnamefont {Samwer}}, \ and\ \bibinfo
  {author} {\bibfnamefont {P.}~\bibnamefont {Vana}},\ }\href@noop {} {\bibfield
   {journal} {\bibinfo  {journal} {Macromol.\ Chem.\ Phys.}\ }\textbf {\bibinfo
  {volume} {215}},\ \bibinfo {pages} {1563} (\bibinfo {year}
  {2014})}\BibitemShut {NoStop}%
\bibitem [{\citenamefont {Deam}\ and\ \citenamefont
  {Edwards}(1976)}]{Deam-Edwards}%
  \BibitemOpen
  \bibfield  {author} {\bibinfo {author} {\bibfnamefont {R.~T.}\ \bibnamefont
  {Deam}}\ and\ \bibinfo {author} {\bibfnamefont {S.~F.}\ \bibnamefont
  {Edwards}},\ }\href@noop {} {\bibfield  {journal} {\bibinfo  {journal}
  {Philos.\ Trans.\ R.\ Soc.\ London Ser.~A}\ }\textbf {\bibinfo {volume}
  {280}},\ \bibinfo {pages} {317} (\bibinfo {year} {1976})}\BibitemShut
  {NoStop}%
\bibitem [{\citenamefont {Gomez}\ \emph {et~al.}(2006)\citenamefont {Gomez},
  \citenamefont {Das}, \citenamefont {Chakraborty}, \citenamefont {Pople},\
  and\ \citenamefont {Balsara}}]{gomez-balsara06}%
  \BibitemOpen
  \bibfield  {author} {\bibinfo {author} {\bibfnamefont {E.~D.}\ \bibnamefont
  {Gomez}}, \bibinfo {author} {\bibfnamefont {J.}~\bibnamefont {Das}}, \bibinfo
  {author} {\bibfnamefont {A.~K.}\ \bibnamefont {Chakraborty}}, \bibinfo
  {author} {\bibfnamefont {J.~A.}\ \bibnamefont {Pople}}, \ and\ \bibinfo
  {author} {\bibfnamefont {N.~P.}\ \bibnamefont {Balsara}},\ }\href@noop {}
  {\bibfield  {journal} {\bibinfo  {journal} {Macromolecules}\ }\textbf
  {\bibinfo {volume} {39}},\ \bibinfo {pages} {4848} (\bibinfo {year}
  {2006})}\BibitemShut {NoStop}%
\bibitem [{\citenamefont {Lay}, \citenamefont {Sommer},\ and\ \citenamefont
  {Blumen}(2000)}]{lay-somm2000}%
  \BibitemOpen
  \bibfield  {author} {\bibinfo {author} {\bibfnamefont {S.}~\bibnamefont
  {Lay}}, \bibinfo {author} {\bibfnamefont {J.-U.}\ \bibnamefont {Sommer}}, \
  and\ \bibinfo {author} {\bibfnamefont {A.}~\bibnamefont {Blumen}},\
  }\href@noop {} {\bibfield  {journal} {\bibinfo  {journal} {J.\ Chem.\ Phys.}\
  }\textbf {\bibinfo {volume} {113}},\ \bibinfo {pages} {11355} (\bibinfo
  {year} {2000})}\BibitemShut {NoStop}%
\bibitem [{\citenamefont {Lay}, \citenamefont {Sommer},\ and\ \citenamefont
  {Blumen}(1999)}]{lay-somm-blum99}%
  \BibitemOpen
  \bibfield  {author} {\bibinfo {author} {\bibfnamefont {S.}~\bibnamefont
  {Lay}}, \bibinfo {author} {\bibfnamefont {J.-U.}\ \bibnamefont {Sommer}}, \
  and\ \bibinfo {author} {\bibfnamefont {A.}~\bibnamefont {Blumen}},\
  }\href@noop {} {\bibfield  {journal} {\bibinfo  {journal} {J.\ Chem.\ Phys.}\
  }\textbf {\bibinfo {volume} {110}},\ \bibinfo {pages} {12173} (\bibinfo
  {year} {1999})}\BibitemShut {NoStop}%
\bibitem [{\citenamefont {Flory}(1941)}]{flory41}%
  \BibitemOpen
  \bibfield  {author} {\bibinfo {author} {\bibfnamefont {P.~J.}\ \bibnamefont
  {Flory}},\ }\href@noop {} {\bibfield  {journal} {\bibinfo  {journal}
  {J.~Am.~Chem.~Soc.}\ }\textbf {\bibinfo {volume} {63}},\ \bibinfo {pages}
  {3083, 3091, 3096} (\bibinfo {year} {1941})}\BibitemShut {NoStop}%
\bibitem [{\citenamefont {Stockmayer}(1943)}]{stockm43}%
  \BibitemOpen
  \bibfield  {author} {\bibinfo {author} {\bibfnamefont {W.~H.}\ \bibnamefont
  {Stockmayer}},\ }\href@noop {} {\bibfield  {journal} {\bibinfo  {journal}
  {J.~Chem.~Phys.}\ }\textbf {\bibinfo {volume} {11}},\ \bibinfo {pages} {45}
  (\bibinfo {year} {1943})}\BibitemShut {NoStop}%
\bibitem [{\citenamefont {Ulrich}(2010)}]{ulrich-diss10}%
  \BibitemOpen
  \bibfield  {author} {\bibinfo {author} {\bibfnamefont {S.}~\bibnamefont
  {Ulrich}},\ }\href@noop {} {Ph.D. thesis},\ \bibinfo  {school} {Univ.\
  G\"ottingen} (\bibinfo {year} {2010})\BibitemShut {NoStop}%
\bibitem [{\citenamefont {Huthmann}\ \emph {et~al.}(1996)\citenamefont
  {Huthmann}, \citenamefont {Rehkopf}, \citenamefont {Zippelius},\ and\
  \citenamefont {Goldbart}}]{endlinks-huthm96}%
  \BibitemOpen
  \bibfield  {author} {\bibinfo {author} {\bibfnamefont {M.}~\bibnamefont
  {Huthmann}}, \bibinfo {author} {\bibfnamefont {M.}~\bibnamefont {Rehkopf}},
  \bibinfo {author} {\bibfnamefont {A.}~\bibnamefont {Zippelius}}, \ and\
  \bibinfo {author} {\bibfnamefont {P.~M.}\ \bibnamefont {Goldbart}},\
  }\href@noop {} {\bibfield  {journal} {\bibinfo  {journal} {Phys.\ Rev.\ E}\
  }\textbf {\bibinfo {volume} {54}},\ \bibinfo {pages} {3943} (\bibinfo {year}
  {1996})}\BibitemShut {NoStop}%
\bibitem [{\citenamefont {Fredrickson}\ and\ \citenamefont
  {Helfand}(1987)}]{fred-helfand87}%
  \BibitemOpen
  \bibfield  {author} {\bibinfo {author} {\bibfnamefont {G.~H.}\ \bibnamefont
  {Fredrickson}}\ and\ \bibinfo {author} {\bibfnamefont {E.}~\bibnamefont
  {Helfand}},\ }\href@noop {} {\bibfield  {journal} {\bibinfo  {journal} {J.\
  Chem.\ Phys.}\ }\textbf {\bibinfo {volume} {87}},\ \bibinfo {pages} {697}
  (\bibinfo {year} {1987})}\BibitemShut {NoStop}%
\bibitem [{\citenamefont {Brazovski\u\i}(1975)}]{brazov75}%
  \BibitemOpen
  \bibfield  {author} {\bibinfo {author} {\bibfnamefont {S.~A.}\ \bibnamefont
  {Brazovski\u\i}},\ }\href@noop {} {\bibfield  {journal} {\bibinfo  {journal}
  {Zh.\ Eksp.\ Teor.\ Fiz.\ [Sov.~Phys.~JETP]}\ }\textbf {\bibinfo {volume} {68
  [41]}},\ \bibinfo {pages} {175 [85]} (\bibinfo {year} {1975})}\BibitemShut
  {NoStop}%
\bibitem [{\citenamefont {Brazovski\u\i}, \citenamefont {Dzyaloshinski\u\i},\
  and\ \citenamefont {Muratov}(1987)}]{braz-dzyal87}%
  \BibitemOpen
  \bibfield  {author} {\bibinfo {author} {\bibfnamefont {S.~A.}\ \bibnamefont
  {Brazovski\u\i}}, \bibinfo {author} {\bibfnamefont {I.~E.}\ \bibnamefont
  {Dzyaloshinski\u\i}}, \ and\ \bibinfo {author} {\bibfnamefont {A.~R.}\
  \bibnamefont {Muratov}},\ }\href@noop {} {\bibfield  {journal} {\bibinfo
  {journal} {Sov.~Phys.~JETP}\ }\textbf {\bibinfo {volume} {66}},\ \bibinfo
  {pages} {625} (\bibinfo {year} {1987})}\BibitemShut {NoStop}%
\bibitem [{\citenamefont {von~der Heydt}, \citenamefont {M\"uller},\ and\
  \citenamefont {Zippelius}(2011)}]{hmz11PRE}%
  \BibitemOpen
  \bibfield  {author} {\bibinfo {author} {\bibfnamefont {A.}~\bibnamefont
  {von~der Heydt}}, \bibinfo {author} {\bibfnamefont {M.}~\bibnamefont
  {M\"uller}}, \ and\ \bibinfo {author} {\bibfnamefont {A.}~\bibnamefont
  {Zippelius}},\ }\href@noop {} {\bibfield  {journal} {\bibinfo  {journal}
  {Phys.\ Rev. E}\ }\textbf {\bibinfo {volume} {83}},\ \bibinfo {pages}
  {051131} (\bibinfo {year} {2011})}\BibitemShut {NoStop}%
\bibitem [{\citenamefont {M\'ezard}, \citenamefont {Parisi},\ and\
  \citenamefont {Virasoro}(1987)}]{mezard}%
  \BibitemOpen
  \bibfield  {author} {\bibinfo {author} {\bibfnamefont {M.}~\bibnamefont
  {M\'ezard}}, \bibinfo {author} {\bibfnamefont {G.}~\bibnamefont {Parisi}}, \
  and\ \bibinfo {author} {\bibfnamefont {M.~A.}\ \bibnamefont {Virasoro}},\
  }\href@noop {} {\emph {\bibinfo {title} {Spin Glass Theory and Beyond}}},\
  \bibinfo {series} {Lecture Notes in Physics}, Vol.~\bibinfo {volume} {9}\
  (\bibinfo  {publisher} {World Scientific},\ \bibinfo {address} {Singapore},\
  \bibinfo {year} {1987})\BibitemShut {NoStop}%
\bibitem [{\citenamefont {von~der Heydt}(2011)}]{alicePhD}%
  \BibitemOpen
  \bibfield  {author} {\bibinfo {author} {\bibfnamefont {A.}~\bibnamefont
  {von~der Heydt}},\ }\href
  {http://webdoc.sub.gwdg.de/diss/2012/vdheydt/vdheydt.pdf} {Ph.D. thesis},\
  \bibinfo  {school} {Univ.\ G\"ottingen} (\bibinfo {year} {2011})\BibitemShut
  {NoStop}%
\end{thebibliography}%

\end{document}